\documentclass[english,prl,aps,twocolumn,floatfix,showpacs,tightenlines,superscriptaddress]{revtex4}
\pdfoutput=1
\usepackage{amsmath,amsfonts,amssymb,latexsym,times}
\usepackage{verbatim}
\usepackage{graphicx}
\usepackage{bm}
\usepackage{epstopdf}

\newcommand{\ket}[1]{| #1 \rangle}

\newcommand{\be}{\begin{equation}}
\newcommand{\ee}{\end{equation}}
\newcommand{\ba}{\begin{eqnarray}}
\newcommand{\ea}{\end{eqnarray}}

\newcommand{\ignore}[1]{}
\def\CC{{\rm\kern.24em \vrule width.04em height1.46ex depth-.07ex
    \kern-.30em C}}
\def\P{{\rm I\kern-.25em P}}
\def\RR{{\rm
         \vrule width.04em height1.58ex depth-.0ex
         \kern-.04em R}}

\def\bbbc{{\mathchoice {\setbox0=\hbox{$\displaystyle\rm C$}\hbox{\hbox
to0pt{\kern0.4\wd0\vrule height0.9\ht0\hss}\box0}}
{\setbox0=\hbox{$\textstyle\rm C$}\hbox{\hbox
to0pt{\kern0.4\wd0\vrule height0.9\ht0\hss}\box0}}
{\setbox0=\hbox{$\scriptstyle\rm C$}\hbox{\hbox
to0pt{\kern0.4\wd0\vrule height0.9\ht0\hss}\box0}}
{\setbox0=\hbox{$\scriptscriptstyle\rm C$}\hbox{\hbox
to0pt{\kern0.4\wd0\vrule height0.9\ht0\hss}\box0}}}}
\def\bbbz{{\mathchoice {\hbox{$\sf\textstyle Z\kern-0.4em Z$}}
{\hbox{$\sf\textstyle Z\kern-0.4em Z$}}
{\hbox{$\sf\scriptstyle Z\kern-0.3em Z$}}
{\hbox{$\sf\scriptscriptstyle Z\kern-0.2em Z$}}}}

\begin{document}
\title{Universal sub-leading terms in ground state fidelity}
\author{Lorenzo Campos Venuti} \email{campos@isi.it}
\affiliation{
Institute for Scientific Interchange, Viale
Settimio Severo 65, I-10133 Torino, Italy
}

\author{Hubert Saleur}
\affiliation{Institut de Physique Th\'eorique CEA, IPhT, CNRS, URA 2306, F-91191 Gif-sur-Yvette, France }
\affiliation{Department of Physics and Astronomy, Center for Quantum Information Science\&Technology, University of Southern California, Los Angeles, CA 90089-0484}
\author{Paolo Zanardi}
\affiliation{Department of Physics and Astronomy, Center for Quantum Information Science\&Technology, University of Southern California, Los Angeles, CA 90089-0484}
\affiliation{
Institute for Scientific Interchange, Viale
Settimio Severo 65, I-10133 Torino, Italy
}

\date{Monday 14, 2008}

\begin{abstract}
The study of the (logarithm of the)  {\em fidelity} i.e., of the  overlap amplitude, between ground states of Hamiltonians corresponding to different coupling constants,  provides a valuable  insight  on critical phenomena. When the parameters are infinitesimally close, it is known that the leading term behaves as
$O(L^\alpha)$ ($L$ system size) where $\alpha$ is equal to the spatial dimension $d$  for gapped systems, and otherwise depends on the critical exponents. 
Here we show that when  parameters are changed along a critical manifold,  a sub-leading
$O(1)$ term   can  appear. This term, somewhat similar  to the topological entanglement entropy, depends only on the system's universality class and encodes non-trivial information about the topology of the system. We relate it to 
universal $g$ factors and partition functions of (boundary) conformal field theory in $d=1$ and $d=2$ dimensions.   Numerical checks  are presented on the
simple example of the XXZ chain.
\end{abstract}


\maketitle
{\em Introduction.---} 
Let $|\Psi(\lambda)\rangle$ denote the ground state (GS) of a system with hamiltonian $H(\lambda)$ depending on a set of parameters $\lambda.$
We define the ground-state fidelity associated to the pair of parameter points $\lambda$ and $\lambda^\prime$
as follows
\begin{equation}
{\cal F}(\lambda,\lambda^\prime):=|\langle\Psi(\lambda)|\Psi(\lambda^\prime)\rangle|\,.
\label{fid}
\end{equation}
This quantity  might provide valuable novel insight
for systems exhibiting quantum phase transitions  \cite{fid-0,zhou,fid-1}, in particular when  there are no obvious local order parameters, but some sort of  
topological order \cite{topo}. 
The strategy advocated in Refs \cite{DG} and \cite{caza}
is differential geometric in nature. The parameters $\lambda$ and $\lambda^\prime$ are chosen infinitesimally close to each other and one focusses on 
the leading term, the fidelity metric or susceptibility $\chi_L$,  in the expansion of (\ref{fid}) as a  function of $\delta\lambda:=\lambda-\lambda^\prime$:
${\cal F} \simeq 1-\delta\lambda^2 \chi_L(\lambda)/2$.
Critical lines can be identified as singular  points of the fidelity metric (in the thermodynamical limit) \cite{DG} or by its 
finite-size scaling \cite{caza}. In particular in \cite{caza} it has been shown that the leading finite-size term in the fidelity metric
is always extensive for gapped systems whereas at the critical points its singular part obeys the  scaling $\chi_L/L^d \sim L^{ 2 z +d -2\Delta_\lambda},$ where $z$ is the dynamical exponent and $\Delta_\lambda$ the scaling dimension of the
operator coupled with $\lambda.$ For sufficiently relevant interactions one sees that the fidelity metric can display a super-extensive behavior
that in turns is responsible for the fidelity drops observed at the quantum phase transition (QPT).  On the other hand for  marginal perturbations  $\Delta_\lambda=d+z$
i.e., when one is moving along a manifold of critical points, the above scaling formula does not provide 
a definite prediction as both log-like terms and $O(1)$ might appear. Accordingly moving along a line of gapless point may not give rise
to a detectable fidelity drop \cite{caza,nodrop}.

In this paper we shall demonstrate that  the finite-size expansion of the GS fidelity $(\ref{fid}),$ when $\lambda$ and $\lambda^\prime$ are 
critical, may feature sub-leading terms of order one that depend only on the universality class of the considered model
and encode non-trivial information about the system topology. This goal will be achieved by establishing 
 connections to (boundary) conformal field theories (BCFT) \cite{bcft,difra}. We shall discuss the 1+1 free-boson case  with the support of  exact diagonalization
results for the critical XXZ chain,  and the 2+1 quantum eight-vertex model \cite{8V}. Finally extensions to the case where one of the parameters corresponds to a gapped phase will be discussed, and  potential connections with entanglement measures proposed.
 
{\em Fidelity and critical theories with boundary.---}
We would like now to establish, on general grounds,  { a connection between the  GS fidelity (\ref{fid}) and the partition function 
of a  classical statistical mechanics system with a boundary interface between regions with different coupling} $\lambda$ and $\lambda^\prime$.
This can be qualitatively understood in terms of the  usual correspondence between wave functions and path integrals.
For the sake of concreteness we will now consider space  to be compactified on a $d$ dimensional hypercube of 
linear size  $L$ (with say periodic boundary conditions)
and, since we are interested in GSs, imaginary time to be  infinitely extended along the $x_\tau$-axis.
The scalar product of GSs then becomes the (properly) normalized partition function for a theory on an infinitely long  cylinder split into two regions 
with different couplings $\lambda$ and $\lambda^\prime$.
More quantitatively, one can prove the following
\begin{equation}
{\cal F}(\lambda,\lambda^\prime) =
\lim_{L_\tau\rightarrow\infty} {Z(\lambda,\lambda^\prime)\over \sqrt {Z(\lambda)Z(\lambda^\prime)}}\,.
\label{Z-fid}
\end{equation}
%
Here $Z(\lambda)$ is the partition function for the homogeneous system of size $2L_\tau \times L$ and $Z(\lambda,\lambda^\prime)$ is the partition function in the same system with one interface.

To convince oneself of the validity of Eq. (\ref{Z-fid}) one can proceed in different ways.
Let us assume, for example that
the underlying classical 2D statistical model, with size $L_\tau\times L$ can be described by a transfer matrix $T(\lambda).$
For $L_\tau \rightarrow \infty$, one has    $\left| \Psi(\lambda) \right \rangle = T(\lambda)^{L_\tau} \left| \Phi \right \rangle / \sqrt{Z(\lambda)}$ with $\left | \Phi \right \rangle$ not orthogonal to the ground state and $Z(\lambda)$ the partition function of a homogeneous system of size $2L_\tau \times L$ and boundary conditions which depend on the quantum model and on $\left | \Phi \right \rangle$.
Then (\ref{Z-fid}) follows immediately, by introducing the spectral resolutions of the $T$'s and taking the imaginary time limit $L_\tau\rightarrow \infty$ to enforce the  projection onto the ground state. Then $Z(\lambda,\lambda^\prime)$ is the partition function for the system with one interface and $Z(\lambda) = Z(\lambda,\lambda)$.


The sort of inhomogeneous system we have in mind is often  better seen as a system with a {\em boundary}. This is easily done 
 by folding: instead of having fields on both sides of the interface (where the scalar product is evaluated) one can consider fields only on the left side with coordinate $x_\tau\leq 0$ and fold the fields living on $x_\tau>0$ into the left domain
 by introducing new species. The problem then becomes a boundary one for a theory with double the number of species, and  some boundary condition (BC) at $x_\tau=0$.

{\em Boundaries and impurities.---}
Let us for the moment focus on  one-dimensional quantum systems $d=1$. By again using the standard mapping to 2D classical system we have  $\ln Z(\lambda,\lambda^\prime)= \ln z_L - L L_\tau f$ where $f$ is a non-universal  bulk term and $\ln z_L$ is a term
associated with the boundary itself. One can now go back to a $d=1$ quantum point of view 
but this time with space along the $x_\tau$ axis, and $L$ interpreted as the inverse temperature $\beta$.  One can write the free energy associated with the boundary as $L f_b:= - \ln z_L=Lu - s$. 
In the critical case for $L\rightarrow\infty$ the latter term  term  gives rise to a degeneracy $O(1)$ factor  $g= e^s$, which is, by scaling, independent of $L$  \cite{affleck}. 
This boundary degeneracy - or equivalently $s$, often  referred to as the boundary entropy -  has played a major role in the analysis of boundary conformal field theories (BCFTs). It has been proven in particular that it is {\em universal}, and thus  depends only on the universality class of the critical theory, and the type of conformal boundary condition \cite{friedan}: for instance, for the Ising universality class with free boundary conditions, $g=1$ , while for fixed boundary conditions $g={1\over\sqrt{2}}$. 
%
 Note that the issue of scalar product of ground states 
occurred in this context very early on, through considerations of the Anderson orthogonality catastrophe \cite{anderson}.


{ \em Fidelity and BCFT.---}
We consider first   the archetypal problem of a two dimensional free boson with two different values of the coupling constants. We write the action as
%
$S=\sum_{i=1}^2 {\lambda_i\over 2}\int_{D_i} (\partial_\mu\varphi_1)^2\label{bosact}
$
%
 where $D_{1,2}=\mathrm{R}^\pm\times [0, L]$.
The only condition we put at the `interface' $x_\tau=0$ is that the fields are continuous (this corresponds to taking the scalar product of wave functions).

We first recall that for a single free boson with coupling $\lambda$, compactified on a circle $\varphi\equiv \varphi+2\pi$, the g factors are $g_D=2^{-1/2}(\pi\lambda)^{-1/4}$ and $g_N=(\pi\lambda)^{1/4}$, for Dirichlet (D) and Neumann (N) boundary conditions respectively. 
Assuming now both bosons compactified on a circle of circumference $2\pi$, and using the equations of motion to fold the system in half gives rise to an equivalent problem of {\em two orthogonal} species of bosons with the same compactification radius, one seeing N boundary conditions with coupling $\lambda_N=\lambda_1+\lambda_2$, the other seeing D boundary conditions with coupling $\lambda_D={\lambda_1\lambda_2\over (\lambda_1+\lambda_2)}$. The total g factor is thus
\begin{eqnarray}
g=g_D(\lambda_D)g_N(\lambda_N)={1\over \sqrt{2}}\left({\lambda_N\over\lambda_D}\right)^{1/4}
=\sqrt{\lambda_1+\lambda_2\over 2\sqrt{\lambda_1\lambda_2}}\label{result}
\end{eqnarray}
Of course we recover that $g=1$ when $\lambda_1=\lambda_2$. Moreover, because one field sees D and the other N, it is clear that in fact the final result does not depend on the compactification radius (and is homogeneous in $\lambda$s).

It is instructive to recover this result via a direct computation (see also \cite{yang}).
First note that, in this non-interacting case, one can show that $Z(\lambda_1,\lambda_2)=Z(\frac{\lambda_1+\lambda_2}{2} )$. Second, since all modes contribute identically to the ratio of partition functions (one is simply dealing with Gaussians), the full fidelity is $F(\lambda_1,\lambda_2)=\prod_{k\neq 0} \sqrt{{2\sqrt{\lambda_1\lambda_2}\over \lambda_1+\lambda_2}}$
where the product is taken over the Brillouin zone. This means $L-1$ modes, --the zero mode is missing-- and thus we have $F= e^{-fL} g$, with $g$ the same as (\ref{result}) of course. While in this calculation $f$, the bulk term, is identical to $\ln g$, we emphasize that, unlike $g$, $f$ is not universal and depends in general on the details of the model. 

\begin{figure}
\begin{centering}
\includegraphics[width=64mm,keepaspectratio]{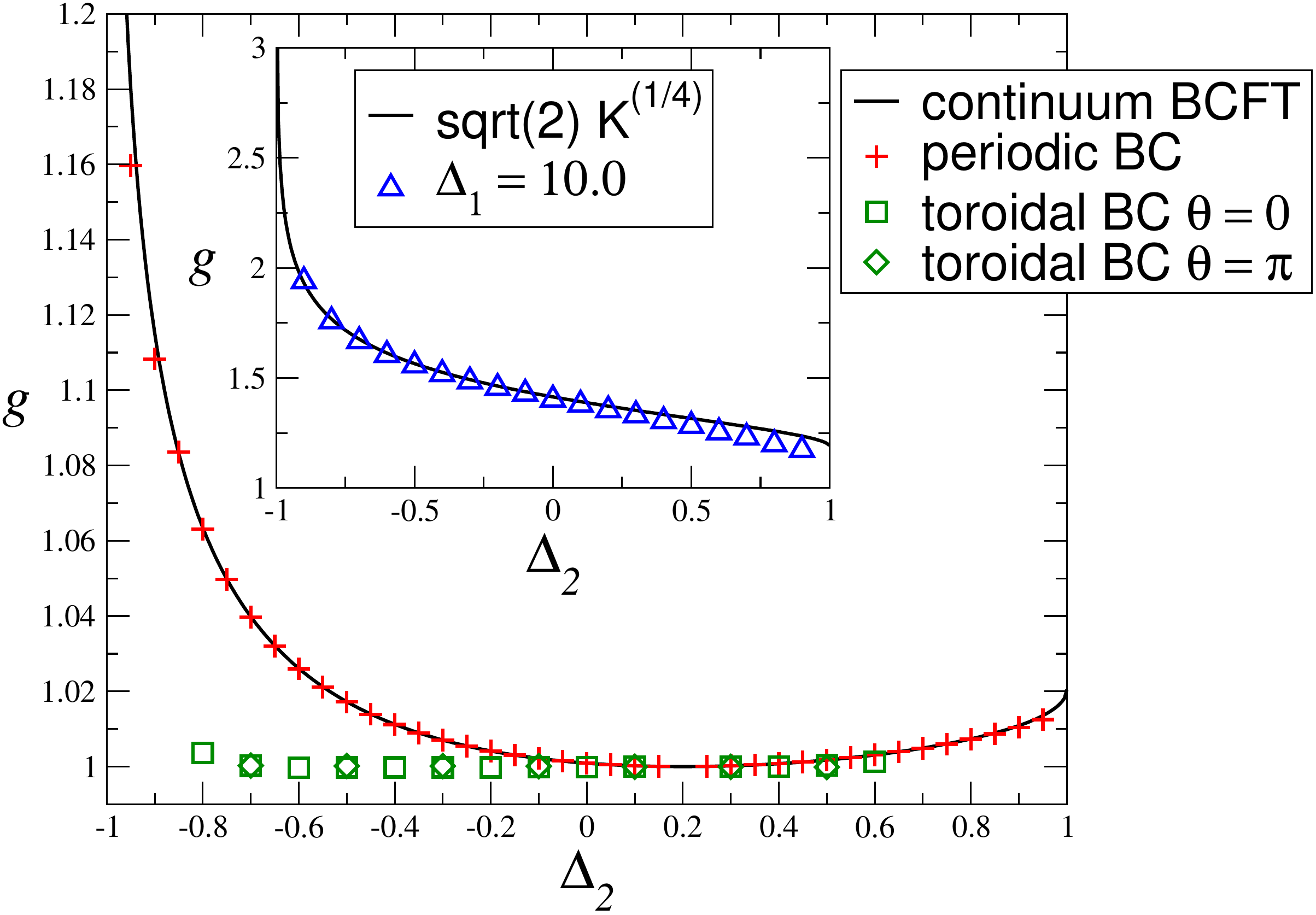}\par
\end{centering}
\caption{(color online). Universal $g$-factor in the fidelity for the $XXZ$ model together with BCFT predictions. The fidelity is computed between ground states with different anisotropies $\Delta_{1,2}$ in the critical region ($|\Delta_{1,2}| < 1$), and we fixed $\Delta_1 = 0.20$. Plus sign are extrapolations of data obtained with Lanczos diagonalization on very small lattices (length $L\le 22$) and periodic boundary conditions. These data agree perfectly with BCFT predictions Eq.~(\ref{result}) together with Eq.~(\ref{eq:K-lambda}). Instead, toroidal BC given by $\sigma^{\pm}_{L+1} = e^{\pm i \theta} \sigma^{\mp}_1$, $\sigma^z_{L+1} = -\sigma^z_1$ induce antiperiodic BC on the field $\phi$ and no $g$ factor (i.e. $g = 1$).  Note that such BC (they belong to conjugacy class (IV) of Ref.~\cite{alcaraz88}) break  conservation of total magnetization.  
In the inset the fidelity is computed when one ground state is critical and the other is deep the massive (N\'{e}el) phase $\Delta_1\gg1$. Solid line give the BCFT prediction. The small discrepancy around $\Delta_2 \approx 1$ is due to finite size effects which are more pronounced near the Kosterlitz-Thouless point $\Delta = 1$. 
\label{fig:XXZ_theo_num}}
\end{figure}

It is interesting to check our prediction against some quick numerical calculations. We thus consider $XXZ$ spin chains defined on a circle of length $L$ with anisotropy $\Delta$. Going over the standard fermionization and bosonization steps \cite{affleckhouches} and matching the results with Bethe-Ansatz, one finds that the continuum limit corresponds to 
\begin{equation}
\lambda= {1\over 2\pi^2}\left[ \pi-\hbox{arccos}\Delta\right]\equiv {1\over 4\pi K} \label{eq:K-lambda}.
\end{equation}
Here we used the conventions where the spin $\sigma^z_i$ of the spin chain is described by $\partial_x\varphi$, and $K$ is an alternative coupling constant often used in the condensed matter literature.
In Fig.~\ref{fig:XXZ_theo_num} we report the results obtained for the lattice $XXZ$ model together with the theoretical predictions based on BCFT Eq.~(\ref{result}) and Eq.~(\ref{eq:K-lambda}); the agreement is very good.
We note that the $g$ factor does depend on  boundary conditions. For instance it is possible, by breaking the $O(2)$ symmetry of the XXZ chain, to 
induce antiperiodic conditions on the fields $\phi$  in the $x$ direction; a quick calculation shows then that  the  term $O(1)$ in the fidelity disappears, i.e.~ $g=1$ in this case, again in agreement with our numerics.

This kind of calculation admits many variants. Instead of having both hamiltonians involved in the fidelity  critical, we can decide to have only one. In this case, the massive  side induces a conformal boundary condition on the critical side in the calculation of $Z(\lambda,\lambda')$, and the term of $O(1)$ in the fidelity is given by the corresponding $g$ factor. We can simulate this situation by turning again to the XXZ model but this time choosing one of the $\Delta$s to be much greater than one.  In this case, the massive side is in the ordered phase, corresponding to two possible ground states, described in terms of spins as $\sigma^z_i=(-1)^i$ and $\sigma_i^z=(-1)^{i+1}$ respectively. Each of these ground states induces Dirichlet boundary conditions for the field $\varphi$ on the massless side. For each of these Dirichlet cases we have $g_D=K^{1/4}$. Meanwhile,  the massive side is a superposition of the two orthogonal ground states with equal coefficients $1\over\sqrt{2}$ so we get in the end $g=2\times{1\over\sqrt{2}}\times K^{1/4}=\sqrt{2}K^{1/4}$. Again these predictions are well confirmed by finite size Lanczos calculation (see inset of Fig.~\ref{fig:XXZ_theo_num}).



{\em Terms of order one in the 2+1 case: the quantum eight-vertex model---}
We turn to consider $O(1)$ terms in the GS fidelity of $2+1$ models whose quantum critical points 
have  dynamical critical exponent  $z=2$.  For these models at criticality, ground state functionals are conformal invariant in the 2d physical space, and equal time correlators coincide with correlations in a 2d CFT. We will now show that the fidelity involves universal terms of $O(1)$ in this case as well, and that this time they are related to partition functions of CFTs on Riemann surfaces. 

To make things concrete, let us specialize to the $2+1$ analog of the free boson - the quantum Lifschitz model - for which a convenient lattice realization is provided by the
 quantum vertex model  \cite{8V}.
The Hilbert space of this  model is spanned by an orthonormal basis  $\{\ket{\mathcal{C}}\}$ in one-to-one correspondence with the configurations  of the classical eight-vertex model. 
The hamiltonian has  the form $H = \sum_{i}Q_i$, with $Q_i$ positive operators, chosen such that $H$ annihilates the following state:
$\ket{\Psi(c^2)}=\sum_{\{\mathcal{C}\}}c^{\hat{n}_c(\mathcal{C})}\ket{\mathcal{C}}/
\sqrt{Z_{2D}(c^2)},\nonumber
$
where  we have chosen, for simplicity, $a=b=1, d=0$ so 
the only remaining parameter is $c$, which is the equivalent here of  $\lambda$ in the previous sections. 
The normalization factor is given by the partition function of the classical eight-vertex model: 
$Z_{2D}(c^2)=\sum_{\{\mathcal{C}\}}c^{2\hat{n}_c(\mathcal{C})}$, 
where $\hat{n}_c(\mathcal{C})$ is the number operators for the $c$type vertices, for the configuration $\mathcal{C}.$ The ground-state phase diagram for the quantum model is identical to the classical one, but given in terms of $c^2.$  The scalar product of ground states is given by
\begin{equation}
\langle\Psi|\Psi^\prime\rangle={Z_{2D}(cc')\over\sqrt{Z_{2D}(c^2)Z_{2D}((c')^2)}}\label{fun}
\end{equation}
The  models live in the 2d equal  time slice, and as usual we will take the dimensions $L_1,L_2$ of this slice to infinity. 
Now  consider the case where the weights obey $c^2\leq 2,(c')^2\leq 2,|cc'|\leq 2$. In that case we are dealing with partition functions of two dimensional critical vertex models, which are described in the continuum limit by Euclidian free bosons in 2d \cite{difra} . With {\em periodic boundary conditions} for instance, these partition functions 
behave as 
%
$Z_{2D}=e^{-fL_1L_2} Z_{CFT}(L_1/L_2)
$
%
where $Z_{CFT}$ is the modular invariant partition function of the conformal invariant field theory. 

\begin{figure}
\begin{centering}
\includegraphics[width=50mm,keepaspectratio]{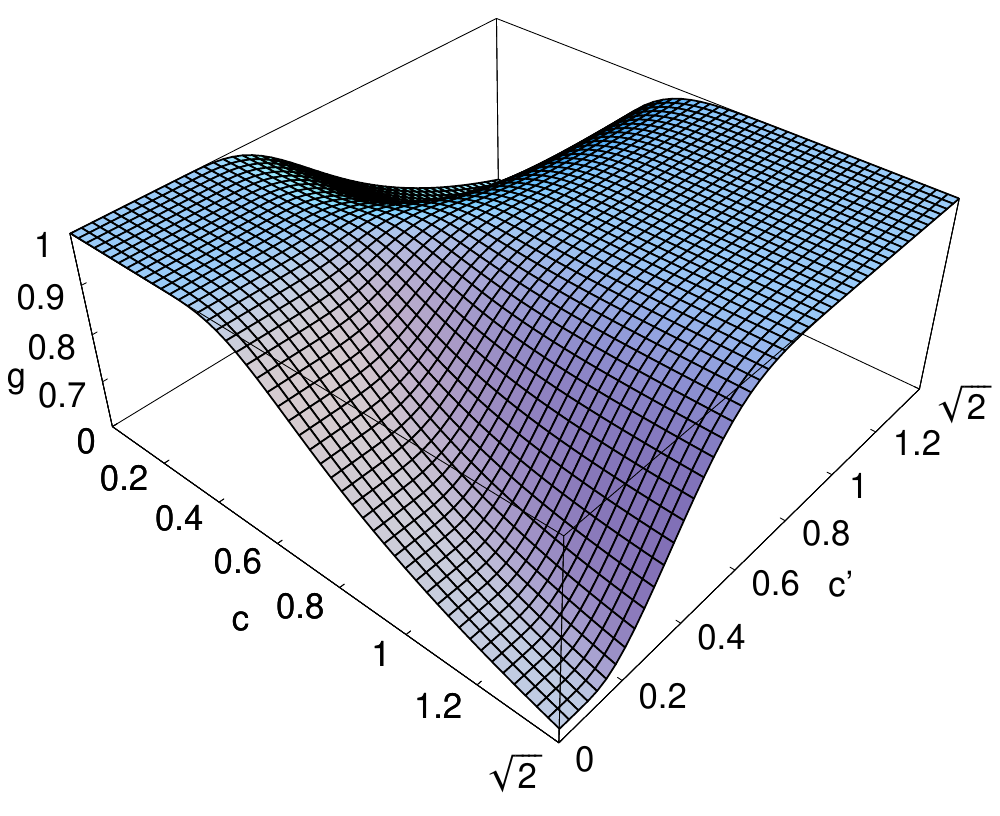}\par
\end{centering}
\caption{(color online). Universal $g$-factor in the fidelity of the quantum eight-vertex model with periodic BC when all the theories are in the disordered region, i.e.~$c^4<4,\,(c')^4<4, \, (c c')^2<4$. We fixed $L_1 = L_2$. The $g$ factor is smooth at the border of the region $c, c' \to 0$ and $c, c' \to \sqrt{2}$.
\label{fig:g2D}}
\end{figure}

%
%
The important point is that the scalar product (\ref{fun}) will have a term behaving like an exponential of the area 
$\exp \left\{-L_1L_2 \left[f(cc^\prime)-f(c^2)/2-f((c^\prime)^2)/2\right]\right\}$ and a term of order one, 
%
${Z(\Lambda) Z^{-1/2}(\lambda)Z^{-1/2}(\lambda^\prime)}
$
%
where $\lambda$ is formally the same  coupling constant  as before (\ref{eq:K-lambda})
 and $\Lambda$ is the coupling associated with the product $cc'$:
%
$\Delta={c^4\over 2}-1=-\cos2\pi^2\lambda\nonumber,
{(c')^4\over 2}-1=-\cos2\pi^2\lambda'\nonumber,
{(cc')^2\over 2}-1=-\cos2\pi^2\Lambda.
$
The conformal partition function itself reads 
$
Z(\lambda)=(\eta(q)\eta(\bar{q}))^{-1} {\cal I}(\lambda)$
where: $q=\bar{q}=\exp(-2\pi L_1/L_2)$ parametrizes the torus,  ${\cal I}(\lambda)=
\sum_{n,m=-\infty}^{\infty}  q^{\frac{1}{4}({\frac{n}{ \sqrt{2\pi\lambda}}+{m}{\sqrt{2\pi\lambda}}})^2}
\bar{q}^{\frac{1}{4}({\frac{n}{ \sqrt{2\pi\lambda}}-{m}{\sqrt{2\pi\lambda}}})^2}
$
and  $\eta(q)=q^{1/24} \prod_{n=1}^\infty (1-q^n)$ \cite{difra}. 
While the prefactor and the $\eta$ terms disappear in the ratio, the instanton sums ${\cal I}$ remain, leading to an rather complicated expression
${\cal I}(cc^\prime)/\sqrt{{\cal I}(c^2){\cal I}((c^\prime)^2)}$  for the  term $O(1)$. 
An example of the behavior of this term is given in figure \ref{fig:g2D}.

Interestingly, being  a combination of conformal partition functions, the term of  $O(1)$ 
 depends heavily on the 
{\em topology} and boundary conditions of the base space.  One can for instance imagine defining the $2d$ quantum models on higher genus Riemann surfaces \cite{Riemann}, or on surfaces with boundaries and  curvature. To give a very simple example, the logarithm of the free-boson partition function  on a rectangle of size $L_1,L_2$ with free boundary conditions (either D on all sides or N on all sides) is given by 
%
$\ln Z_{2D}= aL_1L_2 +b (L_1+L_2)+{1\over 4}\ln L_2
-{1\over 2}\ln\left[ \eta(q)\right]
$
%
where  $a,b$ are non universal terms. The logarithmic term meanwhile is universal, and comes from the general formula for the free energy of a critical region A  of linear size $L_2\propto L_1$,  Euler characteristic $\chi$ and a  boundary with a discrete set of singularities.  Note that all dependence on the coupling constant $\lambda$ has disappeared! It follows that if  we were to calculate the scalar product of ground states in this situation, there would be just no term of order one.


%
%

{\em Connections with  quantum entanglement.---}
Before concluding we would like briefly to comment about the possible connections
between the fidelity approach pursued in this paper and quantum entanglement.
First let us notice that BCFT arguments have been used in the calculations
by Kitaev and Preskill to motivate their expression for the topological entanglement entropy (TEE)
\cite{topo-ent}. Their derivation shows  the  TEE is a O(1) sub-leading universal term 
that is strictly analogous to those for the fidelity in this paper. This is even more manifest
if one expresses the degeneracy $g$ factors in terms of the modular $S$ matrix of the CFT \cite{affleck}
and compares it with the TEE $S_{topo}=\log|S^a_1|.$ 
Moreover there is a  striking similarity between  our  formulas for the log of the fidelity in the previous section and formulas 
in   \cite{FradkinMoore}  for the entanglement entropy at conformal quantum critical points. Both involve logarithms of conformal partition functions, and it is clear that, by taking the ground states of the quantum vertex model with different couplings in different regions, one could obtain entanglement entropy through a term of  $O(1)$ in the fidelity.  How general and useful this observation might be is an open question.

{\em Conclusions.---} Using BCFT techniques, we have shown that the fidelity between critical states contains a term of order $O(1)$ which depends only on the universality class and on the topology of the base space. As such it bears similarity to the topological entanglement entropy or the central charge appearing in the expansion of the ground state energy. The use of methods of CFT in information theory should go much beyond the consideration of these terms of $O(1)$. For example, the same techniques can be used to extract information about the Loschmidt echo \cite{wethesupercool}.

\smallskip
\noindent Acknowledgments: we thank S. Haas and D. Lidar for discussions. 
HS was supported by the ESF Network INSTANS.


\end{document}